\documentclass[10pt,onecolumn,secnumarabic,amssymb, nobibnotes, aps, prd]{revtex4-2}

\usepackage{graphicx}
\usepackage{dcolumn}
\usepackage{bm}
\usepackage{epsfig}
\usepackage{amsmath}

\begin{document}

\title{Discrete ambiguities in a partial-wave analysis of pseudoscalar photoproduction with truncation in total angular momentum}%

\author{A.~Fix}%
\email{fix@tpu.ru}
\affiliation{Tomsk Polytechnic University, 634050 Tomsk, Russia}
\author{I.~Dementjev}%
\affiliation{Tomsk Polytechnic University, 634050 Tomsk, Russia}
\date{}%

\begin{abstract}
The discrete ambiguities appearing in the complete experiment problem for single pseudoscalar meson photoproduction within truncated partial-wave analysis are discussed. It is shown that, in addition to the double ambiguity known from previous works, it is always necessary to take into account another ambiguity arising when truncation in total angular momentum is employed.\\

\noindent PACS: 25.20.Lj, 13.60.Le, 13.88.+e
\end{abstract}

\maketitle

In recent years there has been renewed interest in the problem of a complete experiment for photoproduction of pseudoscalar mesons on nucleons. Due to a large increase in experimental information, it became possible to solve this problem by applying the developed theoretical methods to real data.
In addition to the standard complete experiment problem in terms of spin amplitudes (as a rule, the CGLN amplitudes $F_i$, $i=1,\ldots,4,$ are used), rather powerful methods have been developed within the so-called truncated partial-wave analysis (TPWA) \cite{Grushin,Omelaenko,Wund1}. In this case, the expansion of $F_i$ in terms of partial waves is limited by a certain maximum value $L$ of the angular momentum, and the method is aimed at extracting the coefficients of these series, that is, the multipole amplitudes $E_{l^\pm},M_{l^\pm}$. As observables, one uses the expansion coefficients of spin polarization observables in some complete basis, for example, $\cos^n{\theta}$ or $P_n(\cos{\theta})$, $n=0,1,2,\ldots,$ with $\theta$ being the meson emission angle.

As an important advantage of TPWA over the conventional methods one usually notes simpler requirements to achieve the complete experiment conditions.
In particular, the corresponding complete set of measurements includes, as a rule, a smaller number of double polarization measurements \cite{Grushin,Tiator}.

As applied to photoproduction of pseudoscalar mesons, the most developed approach is based on the Gersten roots method \cite{Gersten} known from the phase analysis of $\pi N$ scattering. The corresponding formalism was mainly developed in \cite{Omelaenko} and then supplemented and extended in \cite{Wund1,Wund2,Wund3}. In the latter works some important aspects related to practical application were also considered in detail.

The mathematical basis of the method is the partial-wave expansion of transversity amplitudes $b_i(\theta)$, $i=1,\ldots,4$, in the form
\begin{subequations}\label{eq1}
\begin{align}
  b_1(\theta) = C\frac{e^{i\theta/2}}{(1+x)^L}\,A_{2L}(x)\,,\quad  b_2(\theta)=-b_1(-\theta)\,,\\
  b_3(\theta) = -C\frac{e^{i\theta/2}}{(1+x)^L}\,B_{2L}(x)\,,\quad b_4(\theta)=-b_3(-\theta)\,,
\end{align}
\end{subequations}
where $x=\tan(\theta/2)$ and $A_{2L}$ and $B_{2L}$ are polynomials of the order $2L$
\begin{equation}\label{eq2}
A_{2L}(x)=\sum\limits_{n=0}^{2L}a_nx^n\,,\quad B_{2L}(x)=\sum\limits_{n=0}^{2L}b_nx^n\,,
\end{equation}
with $L$ being the value of the orbital momentum $l$, at which the multipole expansion is truncated. The polynomial coefficients are linear functions of the multipole amplitudes $E/M_{l^\pm}$. The leading terms and the free terms of $A_{2L}$ and $B_{2L}$ are equal, that is
\begin{equation}\label{eq3}
a_{2L}=b_{2L}\,,\quad a_0=b_0\,.
\end{equation}
Taking into account the relations (\ref{eq3}) the set of $4L$ independent coefficients $a_0,\ldots,a_{2L},b_1,\ldots,b_{2L-1}$ is mapped onto the set of $4L$ amplitudes $E_{l^\pm},M_{l^\pm}$ through a nonsingular linear transformation
\begin{equation}\label{eq4}
(a_n,b_n)\to (E_{l^\pm},M_{l^\pm})\,.
\end{equation}
The constant $C$ in (\ref{eq1}) depends on the definition of the amplitudes $b_i$. Its exact value is irrelevant for further discussion, so it will be omitted in the equations to follow.

The second equality in Eq.\,(\ref{eq3}) leads to the relation between the zeros (Gersten roots \cite{Gersten}) $\alpha_i,\beta_i$ of $A_{2L}$ and $B_{2L}$
\begin{equation}\label{eq5}
\prod\limits_{i=1}^{2L}\alpha_i=\prod\limits_{i=1}^{2L}\beta_i\,.
\end{equation}
Given the complex roots $\alpha_i$ and $\beta_i$, the amplitudes $b_1$ and $b_3$ may be written as (suppressing the irrelevant constant $C$ in (\ref{eq1}))
\begin{subequations}\label{eq6}
\begin{gather}
  b_1(\theta) = \frac{e^{i\theta/2}}{(1+x)^L}\,a_{2L}\prod_{i=1}^{2L}(x-\alpha_i)\,, \\
  b_3(\theta) = -\frac{e^{i\theta/2}}{(1+x)^L}\,a_{2L}\prod_{i=1}^{2L}(x-\beta_i)\,.
\end{gather}
\end{subequations}

\begin{table}[ht]
\centering \caption{Polarization observables in terms of transversity amplitudes, classified into the four groups: $S$ (single polarization), $BT$ (beam-target polarization), $BR$ (beam-recoil polarization), and $TR$ (target-recoil polarization). $I$ is determined as $I=k/q\,d\sigma/d\Omega$, where $k$ and $q$ are the photon and the meson center-of-mass momenta. The shorthand notation $\hat{\cal O}$ stands for the product $I{\cal O}$. The $\pm$ signs in the last column indicate, whether the observable is invariant ($+$), or changes sign ($-$) under the transformations (\ref{eq15}) to (\ref{eq17}).}
\begin{tabular}{@{\hspace{0.7cm}}l@{\hspace{0.9cm}}c@{\hspace{0.8cm}}c@{\hspace{1.0cm}}c@{\hspace{0.9cm}}}
\hline \hline Obs. & Transversity representation & Group\\
\hline
$I$ & $ \frac{1}{2} \left( \left| b_{1} \right|^{2} + \left| b_{2} \right|^{2} + \left| b_{3} \right|^{2} + \left| b_{4} \right|^{2} \right) $ & $S$ & $+$ \\
$ \hat{\Sigma} $ & $ \frac{1}{2} \left( - \left| b_{1} \right|^{2} - \left| b_{2} \right|^{2} + \left| b_{3} \right|^{2} + \left| b_{4} \right|^{2} \right) $ & & $+$ \\
$ \hat{T} $ & $ \frac{1}{2} \left( \left| b_{1} \right|^{2} - \left| b_{2} \right|^{2} - \left| b_{3} \right|^{2} + \left| b_{4} \right|^{2} \right) $ & & $+$ \\
$ \hat{P} $ & $ \frac{1}{2} \left( - \left| b_{1} \right|^{2} + \left| b_{2} \right|^{2} - \left| b_{3} \right|^{2} + \left| b_{4} \right|^{2} \right) $ & & $+$ \\
\hline
$ \hat{G} $ & $ \mathrm{Im} \left[ - b_{1} b_{3}^{\ast} - b_{2} b_{4}^{\ast} \right] $ & $BT$ & $-$ \\
$ \hat{H} $ & $ \mathrm{Re} \left[ b_{1} b_{3}^{\ast} - b_{2} b_{4}^{\ast} \right] $ & & $+$  \\
$ \hat{E} $ & $ - \mathrm{Re} \left[ b_{1} b_{3}^{\ast} + b_{2} b_{4}^{\ast} \right] $ & & $+$ \\
$ \hat{F} $ & $ \mathrm{Im} \left[ b_{1} b_{3}^{\ast} - b_{2} b_{4}^{\ast} \right] $ & & $-$ \\
\hline
$ \hat{O}_{x} $ & $ \mathrm{Re} \left[ - b_{1} b_{4}^{\ast} + b_{2} b_{3}^{\ast} \right] $ & $BR$ & $+$ \\
$ \hat{O}_{z} $ & $  \mathrm{Im} \left[ - b_{1} b_{4}^{\ast} - b_{2} b_{3}^{\ast} \right] $ & & $-$  \\
$ \hat{C}_{x} $ & $  -\mathrm{Im} \left[ b_{1} b_{4}^{\ast} - b_{2} b_{3}^{\ast} \right] $ & & $-$ \\
$ \hat{C}_{z} $ & $  -\mathrm{Re} \left[ b_{1} b_{4}^{\ast} + b_{2} b_{3}^{\ast} \right] $ & & $+$ \\
\hline
$ \hat{T}_{x} $ & $ \mathrm{Re} \left[ - b_{1} b_{2}^{\ast} + b_{3} b_{4}^{\ast} \right] $ & $TR$ & $+$ \\
$ \hat{T}_{z} $ & $ - \mathrm{Im} \left[ b_{1} b_{2}^{\ast} - b_{3} b_{4}^{\ast} \right] $ & & $-$ \\
$ \hat{L}_{x} $ & $ \mathrm{Im} \left[ - b_{1} b_{2}^{\ast} - b_{3} b_{4}^{\ast} \right] $ & & $-$ \\
$ \hat{L}_{z} $ & $ \mathrm{Re} \left[ - b_{1} b_{2}^{\ast} - b_{3} b_{4}^{\ast} \right] $ & & $+$ \\
\hline
\hline
\end{tabular}
\label{tab1}
\end{table}
In Table \ref{tab1}, expressions for the observables are presented in the transversity basis. The complex conjugation of the polynomial coefficients in Eqs.\,(\ref{eq2})
\begin{equation}\label{eq8}
a_n\to a_n^*\,,\quad  b_n\to b_n^*\,,\quad n=0,\ldots,2L\,,
\end{equation}
leads to the transformation of the transversity amplitudes
\begin{subequations}\label{eq9}
\begin{gather}
b_i(\theta)\to e^{i\theta}b^*_i(\theta)\,,\quad i=1,3\,,\\
b_i(\theta)\to e^{-i\theta}b^*_i(\theta)\,,\quad i=2,4\label{eq9a}\,,
\end{gather}
\end{subequations}
which, as may readily be seen from Table \ref{tab1}, leaves the observables of the group $S$ unchanged.

In view of the one-to-one relationship between the coefficients $(a_n,b_n)$ and the multipole amplitudes $(E_{l^\pm},M_{l^\pm})$, the transformation (\ref{eq8}) generate the corresponding transformation
\begin{equation}\label{eq10}
E/M_{l^\pm}\to \tilde{E}/\tilde{M}_{l^\pm}=E/M_{l^\pm}(a^*_n,b^*_n)\,,
\end{equation}
where $E/M_{l^\pm}(a^*_n,b^*_n)$ are the multipoles calculated according to the same rules (\ref{eq4}) but with complex conjugated coefficients $a_n$ and $b_n$.
The invariance with respect to (\ref{eq10}) results in the discrete ambiguity (called double ambiguity in \cite{Grushin,Wund1}) of the group $S$ observables within TPWA.

Instead of the coefficients $a_n,\,b_n$, one can use the roots $\alpha_i,\,\beta_i$, as was done in \cite{Omelaenko,Wund1}. In this case, as follows directly from (\ref{eq6}), the multipole transformation which do not change the group $S$ observables, reads
\begin{equation}\label{eq11}
E/M_{l^\pm}\to \tilde{E}/\tilde{M}_{l^\pm}=E/M_{l^\pm}(\alpha^*_n,\beta^*_n)\,.
\end{equation}
It is evident that (\ref{eq10}) and (\ref{eq11}) give the same set of the transformed amplitudes $\tilde{E}/\tilde{M}_{l^\pm}$ apart from an overall phase. Indeed, from the Vieta's formulas (taking into account (\ref{eq3}))
\begin{subequations}
\begin{gather}
  \sum\limits_{i=1}^{2L}\alpha_i = -\frac{a_{2L-1}}{a_{2L}}\,, \\
  \sum\limits_{i=1}^{2L}\beta_i = -\frac{b_{2L-1}}{a_{2L}}\,, \\
  \sum\limits_{i<j=1}^{2L}\alpha_i\alpha_j = \frac{a_{2L-2}}{a_{2L}}\,, \\
  \sum\limits_{i<j=1}^{2L}\beta_i\beta_j = \frac{b_{2L-2}}{a_{2L}}\,, \\
  \ldots\quad \ldots\quad \ldots\,, \nonumber\\
  \alpha_1\alpha_2\ldots\alpha_{2L} = \beta_1\beta_2\ldots\beta_{2L}= \frac{a_0}{a_{2L}}
\end{gather}
\end{subequations}
immediately follows that $\alpha_i,\beta_i\to \alpha^*_i,\beta^*_i$ leads to $a_n,b_n\to a^*_n,b^*_n$, so that both methods are equivalent.

Note, that the double discrete ambiguity discussed above has global character in the sense that it is present at any energy $W$. In addition, the so-called accidental ambiguities may arise when the equality (\ref{eq5}) still holds if complex conjugation is applied to only some of the roots on the right and the left hand sides. It is clear, however, that such ambiguities, appearing accidentally at isolated energies $W$, cannot generate branches of solutions. Therefore in a real energy-dependent analysis, such 'point-like' degeneracies should not pose principal difficulties. Situations in which they become dangerous are rather exotic  \cite{Wund3}. For this reason, we will focus only on the double discrete ambiguities of the type (\ref{eq10}).

Using the expressions in Table \ref{tab1} one can show
that the ambiguity (\ref{eq10}) (or (\ref{eq11})) can be resolved by measuring one additional observable $\hat{G}$ or $\hat{F}$, or any of the observables from the sets $BR$ and $TR$. Indeed, as may be seen, the transformation (\ref{eq8}) leads to a sign change of $\hat{G}$ and $\hat{F}$. As for observables from the groups $BR$ and $TR$, because of differences in the transformation rules for the pairs $b_1,b_3$ and $b_2,b_4$ (Eqs.\,(\ref{eq9})) they do not have a definite parity with respect to (\ref{eq10}). For example, for $\hat{O}_x$ we will have
\begin{equation}\label{eq12}
\hat{O}_x\to
Re\left[-b_1^*b_4\,e^{2i\theta}+b_2^*b_3\,e^{-2i\theta}\,\right]\ne \pm \hat{O}_x\,.
\end{equation}

Most of the above results have been presented in earlier works \cite{Omelaenko,Wund1,Wund2,Wund3}. Here we would like to consider another type of the double ambiguity, which turns out to be no less significant than the ambiguity associated with the symmetry transformation (\ref{eq8}). It was mentioned in Ref.\,\cite{Keaton}, but has not yet been discussed in detail in the literature. This ambiguity arises when the partial-wave expansion is truncated in the total angular momentum $j=l\pm\frac12$.
To explain its mechanism, it is convenient to use the expansion of the helicity amplitudes in Wigner rotation matrices $d^j_{\lambda\mu}$
\begin{equation}\label{eq13}
H_{\mu\lambda}(\theta)=\sum\limits_j h_{\mu\lambda}^j\,d^j_{\lambda\mu}(\theta)\,,
\end{equation}
where $j$ is the total angular momentum. Following \cite{Walker}, the four independent amplitudes with $\lambda=\frac12,\frac32$ and $\mu=\pm\frac12$ will be numbered by $i=1,\ldots,4$ (see Table I of Ref.\,\cite{Walker}). Taking into account the relationship between the amplitudes $b_i$ and $H_i$
\begin{subequations}
\begin{gather}
  b_1 = \frac12\,[H_1+H_4+i(H_2-H_3)]\,,\\
  b_2 = \frac12\,[H_1+H_4-i(H_2-H_3)]\,,\\
  b_3 = \frac12\,[H_1-H_4-i(H_2+H_3)]\,,\\
  b_4 = \frac12\,[H_1-H_4+i(H_2+H_3)]\,,
\end{gather}
\end{subequations}
it is clear that the replacement
\begin{equation}\label{eq15}
H_1\to H^*_1,\ H_2\to -H^*_2,\
H_3\to -H^*_3,\ H_4\to H^*_4\,,
\end{equation}
leading to
\begin{equation}\label{eq16}
b_i\to b^*_i\,,\quad i=1,\ldots,4\,,
\end{equation}
leaves all observables of the group $S$ invariant. The symmetry (\ref{eq15}) was noted in Ref.\,\cite{Keaton1,Chiang}. Note that in contrast to (\ref{eq9}) all amplitudes $b_i$ in (\ref{eq16}) are transformed according to the same rule.

Using the definition (\ref{eq13}) and taking into account the realness of the Wigner rotation matrices, the transformation (\ref{eq15}) results in the corresponding replacement of the partial amplitudes
\begin{equation}\label{eq17}
h^j_1\to h^{j*}_1,\ h^j_2\to -h^{j*}_2,\
h^j_3\to -h^{j*}_3,\ h^j_4\to h^{j*}_4.
\end{equation}
From this it immediately becomes clear that this ambiguity can be resolved by adding to the group $S$ one of those observables which are not invariant (change sign) under the replacement (\ref{eq15}) (see the last column in Table \ref{tab1}), that is, one of
\begin{equation}\label{eq18}
\hat{G},\,\hat{F}
\end{equation}
from the group $BT$, or one of
\begin{equation}\label{eq19}
\hat{O}_z,\,\hat{C}_x,\,\hat{T}_z,\,\hat{L}_x
\end{equation}
from the groups $BR$ and $TR$.

We emphasize that the symmetry (\ref{eq17}) occurs when the partial-wave expansion is truncated in $j$ and does not occur when one cuts off in $l$. Similarly, the symmetry (\ref{eq10}) arises only when the truncation is in the orbital momentum $l$ and, in the strict sense, does not hold when one cuts off in the total momentum $j$. At the same time, it is clear that as $L=l_\mathrm{max}$, and, accordingly, $J=j_\mathrm{max}$ increase (at fixed $W$), both symmetries should manifest themselves, regardless of the truncation scheme. To illustrate how this comes about, we derive, by analogy with (\ref{eq1})
similar representation of $b_i$, $i=1,\ldots,4$, using the expansion (\ref{eq13}). Omitting here intermediate steps, we present below only the final results. The resulting expressions read (with $x=\tan(\theta/2)$ and suppressing again the constant $C$):
\begin{subequations}\label{eq25}
\begin{gather}
b_1(\theta) = \frac{1}{(1+x^2)^{2J-\frac32}}\,C_{2J}(x)\,,\
b_2(\theta) = -b_1(-\theta)\,, \\
b_3(\theta) = \frac{-1}{(1+x^2)^{2J-\frac32}}\,D_{2J}(x)\,,\
b_4(\theta) = -b_3(-\theta)\,,
\end{gather}
\end{subequations}
where $C_{2J}(x)$ and $D_{2J}(x)$ are polynomials of the order $2J$ with $J=j_\mathrm{max}$. Similarly to (\ref{eq3}), the relations
\begin{equation}\label{eq23}
c_{2J}=d_{2J}\,,\quad c_0=d_0
\end{equation}
for the free and the leading term coefficients hold.

Expressions (\ref{eq25}) can also be obtained directly from (\ref{eq1}) by eliminating the contribution of the multipoles $E_{l^+}$, $M_{l^+}$ with $l=L=l_{max}$. Then the remaining polynomials in $b_1$ and $b_3$ have a constant root $x=-i$. This feature, which is valid for any $J$, was noted in \cite{Wund2} for the simplest particular case $J=1/2$. Taking into account the identity
\begin{equation}\label{eq26}
e^{i\theta/2}(1-ix)=\left(\cos\frac{\theta}{2}\right)^{-1}
\end{equation}
we come to Eqs.\,(\ref{eq25}).

The transformation (\ref{eq15}) (or (\ref{eq17})), leading to (\ref{eq16}) and resulting in the corresponding double ambiguity, means complex conjugation of all coefficients $c_n$ and $d_n$ of the polynomials $C_{2J}$ and $D_{2J}$. As already noted above, all 16 observables, including those from the groups $TR$ and $BR$ have definite parity (are invariant or change sign) with respect to this transformation.

In order to return from (\ref{eq25}) to (\ref{eq1}) we add the contribution of $E_{L^+},\,M_{L^+}$. It is clear that if the expansions (\ref{eq25}) with the chosen $J$ provide the necessary accuracy, then the added multipoles are small. This leads to
\begin{subequations} \label{eq27}
\begin{gather}
b_1(\theta) \to \tilde{b}_1(\theta)=b_1(\theta) + \frac{e^{i\theta/2}}{(1+x^2)^L}\,\tilde{A}_{2L}(x)\,,\\
b_2(\theta) \to \tilde{b}_2=b_2(\theta) - \frac{e^{-i\theta/2}}{(1+x^2)^L}\,\tilde{A}_{2L}(-x)\,,
\end{gather}
\end{subequations}
and to the analogous expressions for $b_3$ and $b_4$. Here the polynomial $\tilde{A}_{2L}(x)$ contains only $E_{L^+}$ and $M_{L^+}$.

The terms proportional to $e^{i\theta/2}$ and $e^{-i\theta/2}$ lead to different transformation rules for $b_1$ and $b_2$ (and, respectively, for $b_3$ and $b_4$) under the complex conjugation of all coefficients on the r.h.s.\ in Eqs.\,(\ref{eq27}), so that the previous symmetric rule (\ref{eq16}) no longer holds. However, since these 'symmetry-breaking' terms contain only the small amplitudes $E_{L^+},\,M_{L^+}$, their contribution is comparable to the error caused by the truncation of the multipole expansion series.

Thus, as one can see, within a given cutoff scheme, one of the two ambiguities is always approximate. At the same time, the corresponding numerical error is proportional to the contribution of the last amplitudes retained in the expansion series.
Therefore, if TPWA is used correctly, that is, if the discarded terms in the expansion are indeed small, then along with the ambiguity (\ref{eq10}) it is also necessary to take into account the ambiguity (\ref{eq17}).

\begin{figure}[t]
\begin{center}
\resizebox{0.55\textwidth}{!}{%
\includegraphics{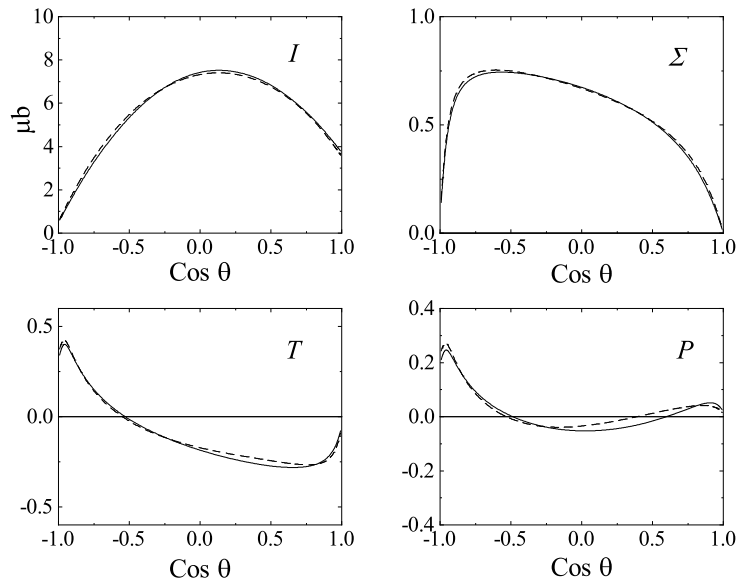}}
\caption{Single polarization observables for $\gamma p\to\pi^0 p$ at $W=1300$ MeV. Calculations are performed with the MAID2007 multipole amplitudes \cite{MAID2007}. Only the $l=0,1,2$ multipoles (from $E_{0^+}$ to $E_{2^+},M_{2^+}$) are included. The solid curves are the starting solution. The dashed curves correspond to the transformation (\ref{eq16}) of the amplitudes $b_i$ on the r.h.s.\ of (\ref{eq27}).}
\label{fig1}
\end{center}
\end{figure}

\begin{figure}[t]
\begin{center}
\resizebox{0.7\textwidth}{!}{%
\includegraphics{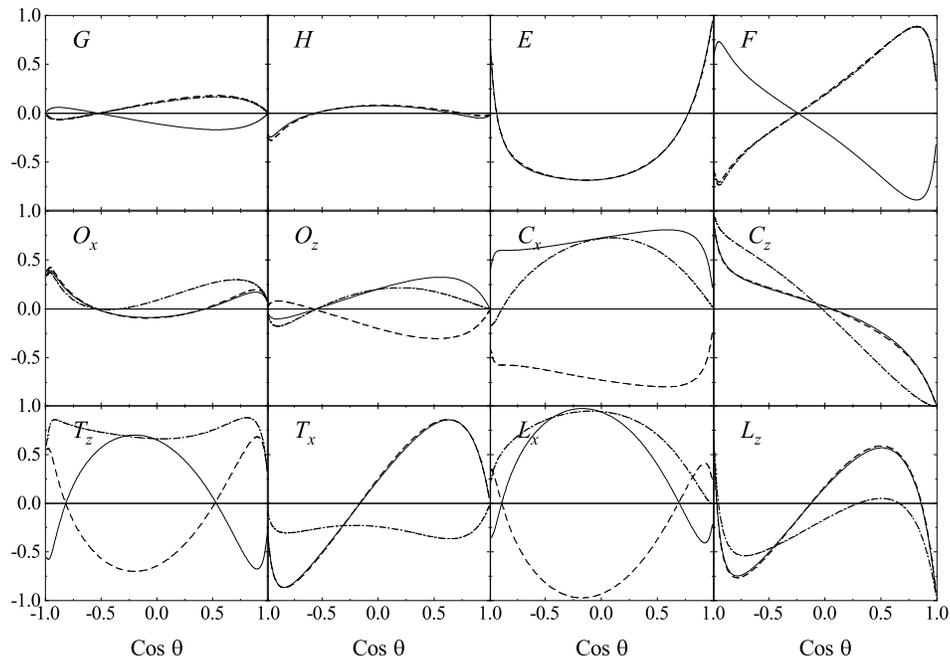}}
\caption{Double polarization observables for $\gamma p\to\pi^0 p$ at $W=1300$ MeV. The meaning of the solid and the dashed curves is as in Fig.\,\ref{fig1}.
Dash-dotted curves are obtained by the transformation (\ref{eq8}). For $H$ and $E$ the dash-dotted curves coincide with the solid curves.}
\label{fig2}
\end{center}
\end{figure}

As an example, Fig.\,\ref{fig1} demonstrates the degree of accuracy, to which the symmetry (\ref{eq17}) holds, if the expansion is truncated in the orbital momentum at $l_\mathrm{max}=L$. Here the group $S$ observables are calculated at $W=1300$ MeV with $L=2$. Since the constraint in orbital momentum is used, the symmetry (\ref{eq8}) is exact, and the observables are invariant under the transformation (\ref{eq10}) with $l=0,\,\ldots,\,2$.

The dashed lines in the same figure are obtained with the amplitudes (\ref{eq27}) in which the first terms are transformed according to (\ref{eq16}), and the second terms proportional $\tilde{A}_{2L}$ are not transformed. Since the symmetry (\ref{eq17}) is violated by the presence of these last terms in (\ref{eq27}), the solutions differ from each other. However, as we can see, this difference caused by the contribution of the small amplitudes $E/M_{2^+}$ is rather insignificant.
This means, in particular, that in a real TPWA with $l\leq L$, along with the additional solution $\tilde{E}/\tilde{M}_{l^\pm}$ (\ref{eq10}), there will be an approximate solution in which the amplitudes $E/M_{L^+}$ do not change.

In Fig.\,\ref{fig2} we also plot the corresponding results for the double polarization observables.
Recall that the observables $\hat{H},\,\hat{E},\,\hat{O}_x,\,\hat{C}_z,\,\hat{T}_x,\,\hat{L}_z$ remain invariant and $\hat{G},\,\hat{F},\,\hat{O}_z,\,\hat{C}_x,\,\hat{T}_z,\,\hat{L}_x$ change sign under the transformation (\ref{eq17}).
The deviation from this rule resulting, in particular, in the difference between the solid and the dashed curves for 
the first six of the above observables
demonstrates the degree of the symmetry breaking, caused by the last terms in (\ref{eq27}).

One important feature must be emphasized. While the ambiguity (\ref{eq10}) may be resolved by adding any observable from the groups $BR$ or $TR$, in the case of (\ref{eq16}), the observables
\begin{equation}\label{eq29}
\hat{O}_x,\,\hat{C}_z,\,\hat{T}_x,\,\hat{L}_z
\end{equation}
belonging to these groups remain invariant. Therefore, in order to eliminate both ambiguities simultaneously, only the observables marked with '$-$' in Table \ref{tab1} should be included into the complete set.

In summary, apart from the double discrete ambiguity, noted in Refs.\,\cite{Grushin,Omelaenko,Wund1,Wund2,Wund3} there is always another ambiguity arising from the invariance of the group $S$ observables under the complex conjugate transformation (\ref{eq17}). The conditions at which these additional uncertainties occur are most naturally controlled by using the partial wave expansion (\ref{eq13}) in the helicity representation.

This result means that the rules for choosing the complete set are more complicated than those given in the above cited works. In particular, both types of the ambiguities can be eliminated simultaneously only by choosing the observables marked by the minus sign in Table \ref{tab1}. On the contrary, taking any observable from (\ref{eq29}) will leave the ambiguity (\ref{eq17}) unresolved.

This work was supported by the Russian Science Foundation, grant No.\,22-42-04401.

\end{document}